\documentclass[aps,prb,twocolumn,superscriptaddress,showpacs,amsmath, amssymb,amsfonts, floatfix]{revtex4}

\usepackage{color}
\usepackage{booktabs}
\usepackage{graphicx,bm}

\newcommand{\nn}{\nonumber}

\newcommand{\be}{\begin{equation}}
\newcommand{\bea}{\begin{eqnarray}}
\newcommand{\ee}{\end{equation}}
\newcommand{\eea}{\end{eqnarray}}

\newcommand{\bc}{\begin{center}}
\newcommand{\ec}{\end{center}}

\newcommand{\ket}[1]{|{#1}\rangle}

\newcommand{\cexp}[1]{\langle{#1}\rangle}

\begin{document}
\title{Quantum Phase Transitions and the Hidden Order in a Two-Chain
Extended Boson Hubbard Model at Half-Odd-Integer Fillings}

\author{Yu-Wen Lee}
\email{ywlee@thu.edu.tw} %
\affiliation{Department of Physics, Tunghai University, Taichung,
Taiwan}

\date{\today}

\begin{abstract}
We study the phase diagram of two weakly coupled one-dimensional
dipolar boson chains at half-odd-integer fillings. We find that
the system contains a rich phase diagram. Four different phases
are found. They are the Mott insulators, the single-particle
resonant superfluid, the paired superfluid, and the bond- or
inter-chain density waves. Moreover, the Mott insulating phase can
be further classified according to a hidden string order
parameter, which is analogous to the one investigated recently in
the one-dimensional boson Mott insulator at integer fillings.
\end{abstract}


\maketitle

\section{Introduction}

The low dimensional electron and spin systems are usually regarded
as important playgrounds for the studies of correlated quantum
matters due to the strong quantum fluctuation effects in one and
two spatial dimensions. Many exotic non-Fermi liquid states such
as the Luttinger liquid and spin liquid states, to name just a
few, are found in the single or coupled chain systems. However,
nature never stops surprising us. The recent advances in the
technique of loading ultracold gases into optical lattices open a
new era in the research of condensed matter systems. Among the
recent achievements, the experimental realization of the
Boson-Hubbard model signatures an important step in this
direction, not only because of the tunability of the controlling
parameters in the corresponding experiments, but also because it
facilitates the first observation of the Mott insulating state of
bosons and the associated quantum phase transition.~\cite{greiner}
While the atomic interactions in the ultracold gases can be
treated as contact ones for most cases, a sizable longer range
interaction is now within experimental reach by using the dipolar
interaction among atoms,~\cite{polar1, polar2, polar3} which could
provide further opportunities of controlling/designing new
experiments. Following these lines of developments, a natural
question to ask is whether or not the longer range interactions
can trigger a stable new quantum phase of matter which contains
non-trivial internal structure.

Among many research works devoted to the understanding of the
effects of long-range dipolar interactions, we mention the recent
work by Emanuele G. Dalla {\it et al},~\cite{Dalla Torre} who
studied the one-dimensional boson insulators within the context of
an extended boson Hubbard model (EBHM) by employing the density
matrix renormalization group (DMRG) method. By tuning the ratio of
the on-site interaction over the hopping amplitude $U/t$ and the
ratio of the longer range interaction over the hopping amplitude
$V/t$, the mean field analysis shows that three different {\it
conventional} phases can be reached. These include the Mott
insulator at large $U$, the density wave state for large $V$ and a
superfluid state for large $t$. The surprising thing is that a new
intermediate insulating state, the Haldane insulator, which
separates itself from the other two insulating states by second
order quantum phase transitions was found in
Ref.~\onlinecite{Dalla Torre}. Moreover, it was shown that such a
state possesses a non-vanishing non-local string order, similar to
the Haldane phase of the quantum spin-one chain. Such a state is
definitely beyond the reach of the traditional one-dimensional
hydrodynamic effective theory for the one-dimensional boson
superfluid-to-Mott transition.~\cite{Giamarchi} Recently, the
present author and his collaborators have developed an
phenomenological {\it two-component} hydrodynamical effective
theory~\cite{ywlee} which successfully captures the main features
of all the phases, including the Haldane insulator, found in the
recent DMRG study of the one-dimensional EBHM at integer fillings.
This effective theory also clarifies the nature of the quantum
phase transitions between different phases.

Knowing the above results, it is desirable to see if similar
exotic phases or insulating states can be found in other
one-dimensional or quasi-one-dimensional systems. As a first
attempt, we consider in this paper a dipolar boson system of two
weakly coupled chains
 within the framework of the EBHM.
Interestingly, we found that the competition of the inter-chain
hopping and the inter-chain interaction does lead to two different
types of Mott insulating states, with one of them possessing a
nontrivial string order. In addition to that, we also found that
the inter-chain attraction can give rise to an interesting {\it
paired} superfluid state where the inter-chain bound boson pairs
show an algebraic long range superfluid order while the
single-boson superfluid correlations decay exponentially. The rest
of the paper is organized as follows: In section II, we introduce
our model and its effective theory. In section III and IV, we
analyze the phases of the model and discuss the issue of the
string order in the Mott insulating state. The final section is
dedicated to our conclusions, and the resulting phase diagram is
summarized in figure 2.

\begin{figure}
\includegraphics[width=2.4in]{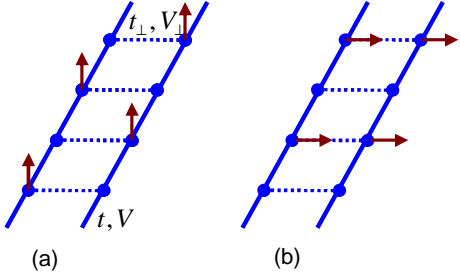}
\caption{ (color online) The two-chain lattice model described by
Hamiltonian (\ref{eqn:bhh}). The arrows denote the possible
orientations of the dipole moments. The left figure (a)
corresponds to the IDW phase and the right figure (b) represents
the BDW phase considered in the paper. (See section III.)}
\label{fig:fig0}
\end{figure}

\section{The model and its effective theory}

The lattice bosons studied in the present paper is described by
the following extended Bose-Hubbard model
\begin{align}\label{eqn:bhh}
H &= -t \sum_{\sigma, \langle i,j \rangle} ( b_{\sigma
i}^{\dag}b_{\sigma j} + \mbox{H.c.}) + \frac{U}{2} \sum_{\sigma i}
\delta n_{\sigma i}^{2} \nn\\& + V \sum_{\sigma, \langle
i,j \rangle} \delta n_{\sigma i}\,\delta n_{\sigma j} \nn\\
&+ t_{\perp}\sum_i(b_{1i}^\dagger b_{2i} + h.c.) +V_{\perp}\sum_i
\delta n_{1i}\delta n_{2i}\, ,
\end{align}
where $\sigma=1, 2$ is the chain index, $t$ is the
nearest-neighbor hopping along the chain, $t_{\perp}$ is the
inter-chain hopping, $U$ is the on-site repulsion, and $V,
V_{\perp}$ are the nearest-neighbor interactions along and between
the chains, respectively. The operator $b_{\sigma i}^\dagger$
create a boson at site $i$ on chain $\sigma$, $n_{\sigma
i}=b_{\sigma i}^\dagger b_{\sigma i}$ is the number operator, and
$\delta n_{\sigma i}=n_{\sigma i}-\bar{n}$ measures the deviation
of the particle number from a mean filling $\bar{n}$. In the
present paper, we will focus our attention to the cases of
half-odd-integer fillings, i.e. $\bar{n}= N+1/2$ where $N$ is a
non-negative integer.

We stress that the above model can be realized by polar molecules,
~\cite{polar1, polar2} or atoms with a larger dipolar magnetic
moment such as $^{53}$Cr.~\cite{polar3} Moreover, it is possible
to adjust the mutual orientations between the dipoles by using
external electric (magnetic) fields, so that we may control the
signs of the nearest-neighbor interactions $V$ and $V_{\perp}$.
Therefore, both the cases of the attractive and the repulsive
nearest-neighbor couplings will be considered below. (See
Fig.~\ref{fig:fig0}.) With this understanding in mind, we now
discuss the low energy effective theory of the model defined in
Eq.~(\ref{eqn:bhh}).


For simplicity, we first consider the case where the on-site
repulsion $U$ is the largest energy scale, i.e. $U\gg t, V,
t_{\perp}, V_{\perp}$. Under this condition and near
half-odd-integer fillings, we can truncate the boson Hilbert space
so that only states with local occupations $n_{\sigma i}=N+1$ and
$N$ are allowed, and the boson operators can be mapped to
spin-$1/2$ operators, $b_{\sigma i}^\dagger \to S_{\sigma i}^+$,
$\delta n_{\sigma i}\to S_{\sigma i}^z$. In this low-energy
subspace, the extended boson Hubbard model is mapped to coupled
$xxz$ spin-$1/2$ models,
\begin{align}\label{spin-ladder}
H&=H_0+H_{\perp} \nn\\
H_0&=-t\sum_{\langle i,j\rangle}(S_{1i}^+S_{1j}^-+
h.c.)+V\sum_{\langle
i,j\rangle} S_{1i}^z S_{1j}^z+(1\to 2)\nn\\
H_\perp&=t_{\perp}\sum_i(S_{1i}^+S_{2i}^-+ h.c.)+V_{\perp}\sum_i
S_{1i}^z S_{2i}^z \, .
\end{align}

In the following, we will assume that the inter-chain coupling
$t_{\perp}, V_{\perp}$ are much smaller than the intra-chain
couplings $t$ and $V$. Then the spin-$1/2$ chain can be bosonized
using the standard method,~\cite{Tsvelik}  and the intra-chain
Hamiltonian can be written in the following form,
\begin{align}\label{intraH}
H_0&=\sum_\sigma \frac{v}{2}\int
dx~\frac{1}{K}(\partial_x\phi_\sigma)^2+K(\partial_x\theta_\sigma)^2\nn\\
&+g\int dx~\cos\sqrt{4\pi}\phi_\sigma\, .
\end{align}
At the perturbative level, $t\gg V$, we have
$v=v_0\sqrt{1+\frac{2V}{\pi t}}\equiv \frac{v_0}{K}, v_0=ta_0$,
and $g=\frac{V}{4\pi^2 a_0}$ ($a_0$ is the lattice spacing.).
However, the validity of this effective action goes beyond its
perturbative derivation, and the relations between the Luttinger
parameter $K$ and $v$ and the spin-chain couplings $t, V$ can be
established exactly, ~\cite{Johnson}
\be\label{Bethe}
v=v_0\frac{\pi\sqrt{1-\Delta^2}}{2\cos^{-1}\Delta},~~
K=\frac{\pi}{2(\pi-\cos^{-1}\Delta)}\, , \ee
with $\Delta=\frac{V}{2t}$ in the above equation.

The inclusion of the inter-chain coupling terms do not present any
new difficulty. Upon including them, the resulting Hamiltonian
becomes
\begin{align}
H&=\sum_{\alpha=s, a}\frac{u_\alpha}{2}\int dx~ \frac{1}{K_\alpha}(\partial\phi_s)^2+ K_\alpha(\partial\theta_s)^2 \nn\\
&+2g \int dx
\cos\sqrt{4\pi}\phi_s\cos\sqrt{4\pi}\phi_a+\frac{t_\perp}{\pi
a}\int dx~\cos\sqrt{2\pi}\theta_a \nn\\
&+\frac{V_\perp}{2\pi a_0}\int
dx~\left(\cos\sqrt{8\pi}\phi_a-\cos\sqrt{8\pi}\phi_s\right).
\end{align}
In the above equation, $\phi_s, \theta_s$ and $\phi_a, \theta_a$
are the symmetric and anti-symmetric combinations of the boson
fields for the spin-$1/2$ operators, respectively. In terms of
them, the original lattice boson operators can be expressed as,
\be\label{single-particle-op} \frac{b_{1/2, i}}{\sqrt{a_0}}
=\frac{1}{\sqrt{2\pi a_0}}
e^{i\sqrt{\frac{\pi}{2}}(\theta_s\pm\theta_a)}\left(1+(-1)^{x/a}\sin\sqrt{2\pi}(\phi_a\pm\phi_a)\right)
\ee
and the bond and inter-chain density fluctuations are
\be\label{density-op} \frac{\delta
n_{s/a}}{a_0}=\sqrt{\frac{2}{\pi}}\partial_x\phi_{s/a}+(-1)^x\frac{-2}{\pi
a_0 }\begin{cases}\sin\sqrt{2\pi}\phi_s\cos\sqrt{2\pi}\phi_a \\
\cos\sqrt{2\pi}\phi_s\sin\sqrt{2\pi}\phi_a
\end{cases}\, .
\ee

Since the intra-chain nonlinear term
$g\cos\sqrt{8\pi}\phi_s\cos\sqrt{8\pi}\phi_a$ is always less
relevant than the inter-chain coupling terms in the region that we
are interested in below, it can be safely neglected. Therefore, in
the strong on-site repulsion limit, the two-chain extended boson
Hubbard model at half-odd-integer fillings can be described by the
following effective hydrodynamic theory,
\begin{align}\label{hydrodynamic}
H&=\sum_{\alpha=s, a}\frac{u_\alpha}{2}\int dx~
\frac{1}{K_\alpha}(\partial\phi_s)^2+ K_\alpha(\partial\theta_s)^2
\nn\\ &+g_1\int dx~\cos\sqrt{2\pi}\theta_a\nn\\&+g_2\int
dx~\cos\sqrt{8\pi}\phi_a+g_3\int dx~\cos\sqrt{8\pi}\phi_s\, ,
\end{align}
where $g_1=\frac{t_\perp}{\pi a_0}, g_2=-g_3=\frac{V_\perp}{2\pi
a_0}$, $v_{s, a}=v \sqrt{1\pm K\frac{V_\perp a_0}{\pi v}}, K_{s,
a}=\frac{K}{\sqrt{1\pm K\frac{V_\perp}{\pi v}}}$, and $v, K$ here
are parameters defined in Eq.~(\ref{Bethe}). At this point, it is
interesting to notice that exactly an action of the same form
occurs in our recent study of the one-dimensional boson Mott
transition near integer fillings,~\cite{ywlee} albeit in a
completely different physical context.


Before we analyze the consequences of the Hamiltonian in
Eq.~(\ref{hydrodynamic}), we emphasize that this effective action
is quite general and is not restricted to hard-core bosons. For
the finite $U$ soft-core bosons, one may introduce the boson field
and density operators,~\cite{Giamarchi}
\begin{align}
\frac{b^\dagger_{i\sigma}}{\sqrt{a_0}}
&=(\rho_0+\frac{1}{\sqrt{\pi}}\partial_x\phi_\sigma)^{1/2}
\sum_p e^{i2p(\pi\rho_0 x+\sqrt{\pi}\phi_\sigma)}e^{i\sqrt{\pi}\theta_\sigma}\nn\\
\frac{n_{i\sigma}}{a_0}&=\left(\rho_0+\frac{1}{\sqrt{\pi}}\partial_x\phi_\sigma\right)\sum_p
e^{i2p(\pi\rho_0 x+\sqrt{\pi}\phi_\sigma)}\, ,
\end{align}
where $\sigma=1, 2$ are the chain indices, and $\rho_0\approx
(N+\frac{1}{2})/a_0$ is the average boson density. After some
simple manipulations, it is not hard to see that we will still get
the same effective action (\ref{hydrodynamic}). Consequently, the
results we get in this paper is robust as long as the inter-chain
couplings are weak. The only problem is that in soft-core boson
case, the relations between the Luttinger parameters $K_{s, a}$
and the microscopic couplings are unknown and Eq.
(\ref{hydrodynamic}) can only be treated in a phenomenological
manner.

\section{The phases and phase transitions of the extended Boson Hubbard model}

After establishing our effective theory for the two-chain EBHM, we
now turn to study the possible phases and the associated quantum
phase transitions of this system. The first thing to notice is
that the charge U(1)$\times$U(1) symmetry of the decoupled chains
is broken down to a U(1)$\times$Z$_2$ by the inter-chain hopping
term $t_\perp (b_{1i}^\dagger b_{2i}+ h.c.)$, which can be viewed
as the inter-chain Josephson coupling between two one-dimensional
boson superfluids. We also notice that the diagonal U(1) symmetry
here forbids terms like $\cos\beta{\theta_s}$. In addition to
that, there is a lattice translation symmetry of the EBHM. As we
shall see immediately, these discrete symmetries can be broken
spontaneously due to correlation effects.

The only nonlinear term which can potentially open a gap in the
symmetric mode is the $g_3$ term in Eq.~(\ref{hydrodynamic}), and
its dimension is $2K_s$. Therefore, we expect that depending on
whether $K_s>1$ or $K_s<1$, the symmetric hydrodynamic mode will
either be gapless or acquire a gap. Furthermore, the transition
corresponding to the opening of the spectral gap in the symmetric
mode falls into the KT universality class. On the other hand, for
the antisymmetric mode, there are two competing nonlinear terms
$g_1$ and $g_2$ in Eq.~(\ref{hydrodynamic}), which have scaling
dimensions $\frac{1}{2K_a}$ and $2K_a$, respectively. Hence,
depending on whether $K_a>1/2$ or $K_a<1/2$, either the
$g_1\cos\sqrt{2\pi}\theta_a$ term or the
$g_2\cos\sqrt{8\pi}\phi_a$ term will be dominant and opens a gap
in the anti-symmetric mode. In other words, the anti-symmetric
mode will always be gapped except at the critical line defined by
$K_a=1/2$, where the critical theory falls into the Ising
universality class. The nature of this transition and the
associated critical mode can be seen as follows:~\cite{Tsvelik}
Exactly at $K_a=1/2$,~\cite{Luther} we may introduce a set of
right- and left-moving Dirac fermions
$\chi_{R,L}=\frac{1}{\sqrt{2\pi a}}e^{\pm\sqrt{4\pi
}\phi^\prime_{R, L}}$ with $\phi^\prime_{R, L}=
\frac{1}{\sqrt{2K_a}}\Phi_a+\mp\sqrt{\frac{K_a}{2}}\Theta_a=\Phi_a+\mp\sqrt{\frac{1}{4}}\Theta_a$
to rewrite the Hamiltonian. In order to identify the critical mode
and the physical low lying excitations, it is convenient to
further decompose the Dirac fermions into their real and imaginary
parts $\chi_{\nu}=\frac{1}{\sqrt{2}}(\xi_\nu+i\rho_\nu), \nu=R,
L$, and express the Hamiltonian in terms of these Majorana
fermions
\begin{align}
H_a&=\frac{-iu_a}{2}(\xi_R\partial_x\xi_R-\xi_L\partial_x\xi_L)+im_\xi \xi_R\xi_L \, , \nn\\
  &+\frac{-iu_a}{2}(\rho_R\partial_x\rho_R-\rho_L\partial_x\rho_L)+im_\rho \rho_R\rho_L \, .
\end{align}
In the above equation, the Majorana fermion masses are
$m_\xi=\frac{g_2+g_3}{4\pi a}, ~m_\rho=\frac{g_2-g_3}{4\pi a}$.
When $|g_2|=|g_3|$, one of the Majorana fermion will become
gapless and the remaining one will still be massive. Hence, we see
that the competition between the inter-chain tunnelling and the
inter-chain interaction may result in an Ising transition. The
above discussions suggest that, depending on $K_s\lessgtr 1$ and
$K_a\lessgtr 1/2$, there can be four different phases. In the
following, we shall elaborate on the natures of these four phases.

We start by considering the case where $K_s>1$ and $K_a>1/2$. In
this case, the only relevant part of the interaction Hamiltonian
is the $g_1$-term, which arises from the inter-chain Josephson
coupling. Using Eq.~(\ref{single-particle-op}) and
Eq.~(\ref{density-op}), one may easily see that the single
particle and density correlations are
\begin{align}\label{SSF}
&\langle b_{1i}^\dagger b_{1j}\rangle=\langle b_{2i}^\dagger
b_{2j}\rangle\sim \frac{1}{|i-j|^{1/4K_s}}\; , \nn\\
&\langle\delta n_{\sigma i} \delta n_{\sigma j} \rangle \sim
\frac{1}{|i-j|^2}+\left(\mbox{\parbox{4cm}{an exponentially decaying\\
staggerred part}}\right)\, .
\end{align}
It is also interesting to examine the following long distance
two-particle correlations,
\begin{align}\label{two-particle1}
\langle b_{1i}^\dagger b_{2i}^\dagger b_{1j}b_{2j}\rangle\sim
\frac{1}{|i-j|^{1/K_s}}\, ,\\
\langle b_{1i}^\dagger b_{2i} b_{1j}^\dagger b_{2j}\rangle\sim
~\mbox{const.} \label{two-particle11}
\end{align}
From the above equations, we see that the single-particle
correlation is dominant over the two-particle pair correlation, as
is expected for a usual superfluid. However,
Eq.~(\ref{two-particle11}) also shows that the system exhibits a
true long-range order for the pair condensate $\langle
b_{1i}^\dagger b_{2i}\rangle\neq 0$. The existence of a non-zero
condensate in this one-dimensional system indicates that some kind
of resonant bond-pairs are formed for bosons on two different
chains, and it is a direct manifestation of the phase-locking
between the two chains due to the inter-chain Josephson coupling.
We also notice that although the density correlations along the
chain decay algebraically, the density-difference $\delta
n_{-i}=:b_{1i}^\dagger b_{1i}-b_{2i}^\dagger b_{2i}:$ has an
exponentially decaying correlation function. This is consistent
with the above resonant bonding-boson pair picture. Since there is
no translational symmetry breaking in this phase, and motivated by
Eq.(\ref{SSF}) and Eq.(\ref{two-particle11}), this phase will be
coined as the single-particle resonant superfluid (RSF) later in
this paper.

We next consider the second gapless phase specified by the
Luttinger parameters $K_s>1$ and $K_a<1/2$. In this case, the
relevant part of the interacting Hamiltonian is $g_2$-term which
originates from the inter-chain dipolar interaction between
bosons. The single particle and density correlations in this phase
are
\begin{align}
&\langle b_{1i}^\dagger b_{1j}\rangle=\langle b_{2i}^\dagger
b_{2j}\rangle\sim \mbox{~decays exponentially~}\, , \nn\\
&\langle\delta n_{\sigma i} \delta n_{\sigma j} \rangle \sim
\frac{1}{|i-j|^2}+(const.)\times\frac{(-1)^{|i-j|}}{|i-j|^{K_s}}\,
.
\end{align}
The most interesting feature of this phase is that although the
symmetric mode is gapless and possess a non-zero superfluid
stiffness, the single-particle superfluid correlation decays
exponentially. On the other hand, two-particle correlation shows
an algebraic long range order,
\begin{align}\label{two-particle2}
&\langle b_{1i}^\dagger b_{2i}^\dagger b_{1j}b_{2j}\rangle\sim
\frac{1}{|i-j|^{1/K_s}}\, ,\nn\\
&\langle b_{1i}^\dagger b_{2i} b_{1j}^\dagger b_{2j}\rangle\sim
~\mbox{~decays exponentially~}\, .
\end{align}
From the above results, we conclude that this translationally
invariant phase is a one-dimensional {\it paired superfluid}
(PSF), where the inter-chain bond-boson pairs flow coherently
along the chain.

We now turn to discuss the remaining two gapped phases. For
$K_s<1$ and $K_a>1/2$, both the inter-chain Josephson and the
inter-chain dipolar interaction are relevant, and these
interactions open gapes for both the symmetric and ant-symmetric
modes. The relevant part of the effective Hamiltonian in the
present case is,
\begin{align}
H&=\sum_{\alpha=s, a}\frac{v_\alpha}{2}\int dx
\left[\frac{1}{K_\alpha}(\partial_x\phi_\alpha)^2+K_s(\partial_x\theta_\alpha)^2\right]\nn\\
&+g_1\int dx~\cos\sqrt{2\pi}\theta_a+g_3\int
dx~\cos\sqrt{8\pi}\phi_s\, .
\end{align}
In this phase, it is easy to see that among the correlation
functions examined previously, the only one which does not decay
exponentially is the resonant-pair condensate Eq.
(\ref{two-particle11}). This is a translationally invariant
insulating state with a non-zero resonant pair boson condensate.
Therefore, it is a Mott insulator (MI).

The last phase corresponds to $K_s<1, K_a<1/2$, and the relevant
part of the effective Hamiltonian in this case is,
\begin{align}
H&=\sum_{\alpha=s, a}\frac{v_\alpha}{2}\int dx
\left[\frac{1}{K_\alpha}(\partial_x\phi_\alpha)^2+K_s(\partial_x\theta_\alpha)^2\right]\nn\\
&+g_2\int dx~\cos\sqrt{8\pi}\phi_a+g_3\int
dx~\cos\sqrt{8\pi}\phi_s\, ,
\end{align}
with $-g_3=g_2\propto V_\perp$ in the simplest EBHM. In this
phase, all the single-particle and the pair correlations decay
exponentially. On the other hand, for the bond- and inter-chain
density fluctuations, $\delta n_{\pm i}\propto:b_{1i}^\dagger
b_{1i}\pm b_{2i}^\dagger b_{2i}:$, we have
\begin{align}
\begin{cases} V_\perp>0, & \langle\delta n_{-i}\rangle\sim
(-1)^{x_i}/a_0,\ , \\
V_\perp<0, & \langle\delta n_{+i}\rangle\sim  (-1)^{x_i}/a_0
\end{cases}\, .
\end{align}
This result indicates that the discrete Z$_2$ lattice translation
symmetry is broken spontaneously. Therefore, depending on the sign
of the inter-chain dipolar interaction, the system exhibits either
a bond-density wave pattern (BDW) or an inter-chain density wave
pattern (IDW). (See Fig.1 for a pictorial illustration.) Notice
that if we change the relative orientations of the dipoles between
the two chains so that the inter-chain dipolar interaction changes
its sign, we will tune a quantum phase transition across these two
different density wave phases and the associated critical theory
is a U(1)$\times$U(1) Gaussian theory. These translationally
non-invariant states can, in principle, be observed experimentally
in the corresponding spatial image patterns of the time-of-flight
experiments.

\begin{figure}
\includegraphics[width=2.4in]{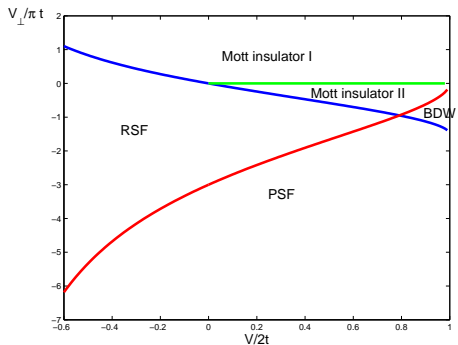}
\caption{ (color online) The phase diagram of the two weakly
coupled half-filled one-dimensional EBHM obtained in the hard-core
boson limit. Notice that for the hard-core bosons, only the BDW
phase can occur. However, we expect that this is a artifact of the
hard-core condition, and for more general cases, depending on the
sign of $V_\perp$, both the IDW and the BDW phases can be
stablized. See the discussions in the last section. Notice also
that the phase transition across the red line is of the Ising
universality class, while the phase transition across the blue
line belongs to the KT universality class. The Mott insulator II
below the green line contains a non-trivial string order, and it
is separated from the Mott insulating I state by a U(1) Gaussian
critical theory. } \label{fig:fig1}
\end{figure}

\section{The hidden string order}

Inside the Mott insulator phase, we can also adjust the relative
orientations of the dipoles between the two chains to change the
sign of $V_\perp$. This will introduce a further quantum phase
transition due to the accidental gap closing in the symmetric mode
alone, and the corresponding critical theory is a U(1) Gaussian
theory. On both sides of the transition, the system is gapped and
do not break any lattice symmetry. So, what is the difference
between these two Mott insulating states? Interestingly, from the
structure of the above hydrodynamic theory and the results of
recent studies on the 1D boson Mott insulator, ~\cite{ywlee,Dalla
Torre} we expect that one of the Mott insulating phase will posses
a non-vanishing non-local string order, ~\cite{commentDW} while
the other does not. This can also be seen by noticing  that our
effective Hamiltonian bears some similarities with that of {\it
isotropic} spin ladders,~\cite{Shelton,Solyom} where nontrivial
string orders were also found. More specifically, drawing the
analogy with the isotropic spin ladders, we may introduce the
string operator
\be \hat{O}_{string}(i-j)=\lim_{|x-y|\to \infty}\cexp{\delta
n_{+i}e^{i\pi\sum_{l=i+1}^{j-1}\delta n_{+l}}\delta n_{+j}}\, .
\label{string-op} \ee
Following of the discussions of M. Nakamura,~\cite{string-order}
we expect that the correct bosonized form of the string operator
should be
\be \hat{O}_{string}\sim\lim_{|x-y|\to\infty}\left\langle\sin
\sqrt{2\pi}\phi_s(x)\sin \sqrt{2\pi}\phi_s(y)\right\rangle \, ,
\ee
which is nonzero for $V_\perp <0$, due to the pining of the
symmetric mode at $\phi_s=\sqrt{\frac{\pi}{8}}$. On the other
hand, the string order vanishes for $V_\perp >0$ where the field
$\phi_s$ is pinned at $\phi_s=0$. Therefore, for the negative
$V_\perp$ case, the MI phase we found here is the boson analogy of
the AKLT phase of a spin ladder system, while for the positive
$V_\perp$ case, the boson MI in the present EBHM is the boson
analogy of the RVB phase of the spin ladder.~\cite{string-order}

A more interesting question is perhaps about how to detect the
differences between the two Mott insulating states. From the
structure of the effective theory, the bulk spectrums of these two
insulating states in the immediate neighborhoods of the transition
line are almost identical. On the other hand, it can be shown that
the MI state with a non-vanishing string order contains a zero
energy edge state.~\cite{edge1} This fact is most easily seen by
examining the Sine-Gordon model of the symmetric mode near its
Luther-Emery point~\cite{Luther} ($K_s=1/2$). Near the
Luther-Emergy point, the symmetric mode can be mapped to a massive
Dirac fermion theory,~\cite{Klein}
\begin{align}
H&=\int_0^\infty dx \Psi^\dagger(x)
\left[-iv\sigma^3\partial_x+\sigma^2 m \right]\Psi(x)+\cdots\, ,
\end{align}
where $\Psi(x)=(\psi_R(x), \psi_L(x))^T$ is a Dirac fermion field,
$m=-V_\perp/2$, and $\cdots$ denotes the possible four fermion
interactions. These fermions should be identified as the low
energy elementary (soliton) excitations of the symmetric mode in
the insulating states. For a system with a boundary at, say $x=0$,
we may introduce the boundary condition
$\psi_R(0)=\psi_L(0)$,~\cite{Boundary} and reformulate the above
Dirac theory in terms of a single chiral component by introducing
$\psi_R(x)=\psi_L(-x)$ for $x<0$. After doing so, the Dirac
Hamiltonian reduces to
\begin{align}
H=\int_{-\infty}^\infty dx
-iv\psi_R(x)^\dagger\partial_x\psi_R(x)+im{\rm
sgn}(x)\psi_R^\dagger(x)\psi_R(-x)\, .
\end{align}
It is not hard to see that the corresponding Schr\"{o}dinger
equation
\be -iv\partial_x\psi_R(x)+im{\rm sgn}(x)\psi_R(-x)=\epsilon
\psi_R(x)\, , \ee
contains a zero energy state localized at the edge,
\be \psi_{Re}(x)\sim e^{-m|x|}\, , ~\epsilon_e=0\, . \ee
Notice that this edge state exists only for $m>0$, i.e. for
$V_\perp<0$. Since all other bulk excitations have finite gapes,
the existence of this zero-energy edge state constitutes a unique
feature of the string-ordered MI state, and its existence can be
revealed in a zero frequency delta-function like peak in the
optical absorption spectrum. Since in real experiments, the system
is trapped in a finite domain, it is interesting to see if such a
delta-function absorption peak can be observed experimentally.

\section{Conclusions and discussions}

Although most of our results are obtained in the weak inter-chain
coupling limit, we point out here that some of the important
features of the weak coupling phase diagrams can be realized even
in the strong inter-chain coupling limit. For example, the effect
of a large $V_\perp$ can be understood by considering the
hard-core boson limit and keeping only the states with local
occupations $n_{\sigma i}=0, 1$.  Under this condition, we may
introduce the bond operators,~\cite{bond-operator}
\begin{align}
s^\dagger\ket{0}&=\frac{1}{\sqrt{2}}(\ket{10}-\ket{01})\, ,
~~t_x^\dagger\ket{0}=\frac{-1}{\sqrt{2}}(\ket{11}-\ket{00})\, ,\nn\\
t_y^\dagger\ket{0}&=\frac{i}{\sqrt{2}}(\ket{11}+\ket{00})\, ,
~~t_z^\dagger\ket{0}=\frac{1}{\sqrt{2}}(\ket{10}+\ket{01})\, ,
\end{align}
for each site, and the ket, say $\ket{10}$, denotes a state with a
local bond-boson occupation $\ket{n_{1i}=1, n_{2i}=0}$, and these
operators obey the constraint $s_i^\dagger
s_i+\sum_{\alpha}t_{\alpha i}^\dagger t_{\alpha i}=1$. In terms of
these bond operators, the inter-chain interaction can be written
as
\be H_{V_\perp}=V_\perp \sum_i (t_{xi}^\dagger
t_{xi}+t_{yi}^\dagger t_{yi})-(s_i^\dagger s_i + t_{zi}^\dagger
t_{zi})\, . \ee
When $|V_\perp|\gg |t|, |t_\perp|$, and $V_\perp >0$, we may keep
only $s_i, t_{zi}$, and the leading order effective low energy
Hamltonian becomes
\begin{align}
H&=-V_\perp +\frac{V}{2}\sum_i (n_{Ai}n_{A i+1}+n_{Bi}n_{B
i+1})\nn\\ &-\frac{V}{2}\sum_i (n_{Ai}n_{B i+1}+n_{Bi}n_{A
i+1})+O(\frac{t^2, t^2_\perp}{V_\perp})\, ,
\end{align}
where $A_i^\dagger=\frac{1}{\sqrt{2}}(s_i^\dagger+t_{zi}^\dagger),
B_i^\dagger=\frac{1}{\sqrt{2}}(s_i^\dagger-t_{zi}^\dagger)$ create
bond-pairs $\ket{10}_i$ and $\ket{01}_i$, respectively, and $n_{A,
Bi}=A_i^\dagger A_i, B_i^\dagger B_i$ are the number operators of
the $A$ and $B$ bosons, respectively. We see from this strong
coupling analysis that in the large $V_\perp$ limit, the ground
state configuration is completely determined by minimizing the
inter-particle interactions between the $A$ and $B$ particles.
Therefore, for $V>0$, we expect a Wigner crystal like state with
the $A$ and $B$ particles are arranged in an alternating pattern
$\cdots ABAB\cdots$, while for $V<0$, phase separation will occur.
In terms of the original two-chain boson language, this Wigner
crystal like density wave state is exactly the IDW phase discussed
previously. When $V_\perp <0$, we get similar results, except now
that the $A$ and $B$ operators are defined by
$A_i^\dagger=\frac{1}{\sqrt{2}}(t_{xi}^\dagger-it_{yi}^\dagger),
B_i^\dagger=\frac{1}{\sqrt{2}}(t_{xi}^\dagger+it_{yi}^\dagger)$,
which create states with local-bond occupations $\ket{11}_i$ and
$\ket{00}_i$, respectively. Again, for $V>0$ the ground state is a
density wave state and for $V<0$ we have phase separations. It is
not hard to see that the Wigner crystal like state in this
situation is just the BDW phase discussed earlier.

Similarly, when the inter-chain hopping amplitude $t_\perp$ is the
largest energy scale, we may introduce the inter-chain bonding and
anti-bonding bosons fields $b_{\pm,
i}=\frac{1}{\sqrt{2}}(b_{1i}\pm b_{2i})$. At low energy, the
anti-bonding bosons can be projected out and the resulting low
energy Hamiltonian becomes
\begin{align}
H_{eff}&=-t_\perp \sum_i b_{+i}^\dagger b_{+i} -t\sum_i
(b_{+i}^\dagger b_{+, i+1}+ h.c.)\nn\\&+\frac{U}{4}\sum_i
(:n_{b_+i}:)^2+\frac{V}{2}\sum_{\langle i,
j\rangle}:n_{b_+i}::n_{b_+j}:\,\nn\\& + O(\frac{U^2, V_\perp^2,
V_\perp^2}{t_\perp})\, ,
\end{align}
to the leading order, where $n_{b_+i}=b_{+i}^\dagger b_{+i}$ is
the density operator of the bonding bosons. This effective
Hamiltonian is nothing but that of the one-dimensional EBHM of the
bonding-bosons at integer fillings. Transforming back to the
original two-chain boson language, it is not hard to see that the
superfluid phase of the bonding boson at large $t$, where the
boson density distributed uniformly, automatically possesses a
non-vanishing resonant-pair condensate $\cexp{b_{1i}^\dagger
b_{2i}+h.c.}$. Hence this phase may be identified as the RSF phase
of the two-chain EBHM considered in this paper. Similarly, its
density wave phase at large $V$ can be viewed as the BDW phase of
the original two-chain system. According to Ref.~\onlinecite{Dalla
Torre}, when $V$ increases to the order of $U$, we expect the
bonding boson will enter into a ``Haldane insulator" phase which
has a non-trivial string order. Since in the truncated Hilbert
space where the anti-bonding bosons are projected out, the density
operator $n_{b_+i}$ coincides with the total bond boson number
operator $n_{+i}$, Hence, the string operator defined in Eq.
(\ref{string-op}) reduces to that of Ref.~\onlinecite{Dalla Torre}
in the large $t_\perp$ limit. Although the transition between the
Haldane insulating phase and the conventional Mott insulating
phase of the bonding-bosons is achieved by tuning the intra-chain
coupling $V$, while it is the inter-chain coupling $V_\perp$ which
plays a decisive role in the corresponding phase transition of the
weakly coupled chain case, the fact that the string operators
coincide in certain limit suggests that these two phases are, in
fact, adiabatically connected.


We now briefly discuss some earlier results which are related to
the present work. A phase analogous to our PSF state was noticed
in the context of a coupled spin model,~\cite{Orignac1} and it was
also shown to appear in a boson ladder model at incommensurate
fillings.~\cite{Orignac2} The Mott-superfluid transition for a
boson ladder model, coupled by the inter-chain josephson coupling
alone, at the commensurate filling of one boson per site was
studied in Ref.~\onlinecite{Donohue}, while in our present case,
the Mott insulating state occurs at half-odd-integer fillings. We
also notice that the hidden order in single chain electronic
Hubbard models was studied using both the bosonization and the
DMRG methods in Ref.~\onlinecite{Nussinov}, and there are some
recent efforts which address the issue of string orders in
electronic band insulators and their relations with the Haldane
chain.~\cite{Rosch1} These works, together with the results of
Ref.~\onlinecite{Dalla Torre} and our work, suggest that the idea
of classifying the boson or fermion Mott insulators in
quasi-one-dimensional systems using the string order may be guite
general.

To summarize, the above strong coupling analysis shows that the
phases we obtained, when one of the inter-chain coupling is
dominant, is consistent with the results we gained in the
weak-coupling bosonization analysis. Since one of the most
important conclusion that follows from the bosonization study is
that the interplay between the inter-chain hopping and inter-chain
interaction results in two exotic phases
--- the paired superfluid state and the Mott insulating state with
a string order, it is tempting to speculate that the major results
of the bosonization analysis in this paper should be valid even
when the inter-chain coupling strength is not so weak and can be
extended to larger regions in the whole phase diagram.

Using the analytic results obtained in this paper as a guide, we
hope that it is helpful for future numerical works to determine
the exact phase boundaries of the two-chain EBHM, and more
importantly, the exotic insulating phase and the paired superfluid
phase can be observed in the future experiments.

\begin{acknowledgments}
The author would like to thank Prof. Yu-Li Lee for critical
suggestions. This research work was support by the National
Science Council of Taiwan under the contract NSC
96-2112-M-029-006-MY3.

\end{acknowledgments}



\begin{thebibliography}{21}
\bibitem{commentDW}
Following this reasoning, one should expect that the non-local
string order will also exist in one of the density wave phase.
This is indeed the case. However, this order does not play any
important role there because the density wave phases already
distinguish themselves from other phases through local order
pamameters.
\bibitem{Klein}
Away from the Luther-Emery point, we expect that the mass of the
Dirac fermions will be renormalized by the current-current
interactions.~\cite{Shelton} However, as long as we are not too
far away from the Luther-Emery point, the existence of the edge
state should not be affected.
\end{thebibliography}
\end{document}